\documentclass[prb,amsmath,amssymb,amsfonts,superscriptaddress,floatfix,showpacs]{revtex4-1}

\usepackage{graphicx}
\usepackage{subfigure}
\usepackage{dcolumn}
\usepackage{bm}
\usepackage{epstopdf}

\begin{document}
\title{A microscopic model for ultrafast remagnetization dynamics}

\author{Raghuveer Chimata}
\affiliation{Department of Physics and Astronomy, Uppsala University, Box 516,
 751\,20 Uppsala, Sweden}
\author{Anders Bergman}
\affiliation{Department of Physics and Astronomy, Uppsala University, Box 516,
 751\,20 Uppsala, Sweden}
\author{Lars Bergqvist}
\affiliation{Department of Materials Science and Engineering, KTH Royal institute of Technology, Brinellv. 23, Stockholm, Sweden}
\author{Biplab Sanyal}
\email{Biplab.Sanyal@physics.uu.se}
\affiliation{Department of Physics and Astronomy, Uppsala University, Box 516,
 751\,20 Uppsala, Sweden}
 \author{Olle Eriksson}
 \affiliation{Department of Physics and Astronomy, Uppsala University, Box 516,
 751\,20 Uppsala, Sweden}

\date{\today }

\begin{abstract}
In this letter, we provide a microscopic model for the ultrafast remagnetization of atomic moments already quenched above Stoner-Curie temperature by a strong laser fluence. Combining first principles density functional theory, atomistic spin dynamics utilizing the Landau-Lifshitz-Gilbert equation and a three temperature model, we show the temporal evolution of atomic moments as well as the macroscopic magnetization of bcc Fe and hcp Co covering a broad time scale, ranging from  femtoseconds to picoseconds.  Our simulations show a variety of complex temporal behavior of the magnetic properties resulting from an interplay between electron, spin and lattice subsystems, which causes an intricate time evolution of the atomic moment, where longitudinal and transversal fluctuations result in a macro spin moment that evolves non-monotonically.
\end{abstract}
\maketitle

Magnetization dynamics of extreme speed has been demonstrated in several experimental reports, starting with the work of laser induced femtosecond dynamics by Beaurepaire {\it et al.} \cite{beaurepaire} Several interesting aspects and questions arise in such studies, where from a technological point of view the reading and writing of information in a magnetic medium stands out as the most interesting. \cite{vahaplar} In the investigation of Ref.~\onlinecite{vahaplar}, a polarized 100 fs laser pulse was used to study switching times of the order of 30 ps. Of more fundamental interest is the question about angular momentum dissipation in ultrafast demagnetization experiments\cite{malinowski,bigot} as well as the different dynamical behavior of spin and orbital angular momenta during a ultrafast demagnetization process. \cite{stamm,boeglin}  

On the theoretical side it has been suggested that different magnetic responses can be found in so called type I and type II ferromagnets, \cite{koopmans} in which the former has a stronger magnetic coupling where spin scattering generates a fast equilibration of the spin and electron systems. For the type II ferromagnets, where a fast demagnetization in the
first picoseconds is followed by a slower
demagnetization, it is suggested that the temperature of the electron system is different 
for the first picoseconds compared to the latter. Theoretical calculations have also been done assuming a two-temperature model, which is coupled to macro spin simulations. \cite{atxitia} 
One of the more discussed aspects of femtosecond dynamics is the dissipation of angular momentum during the demagnetization process. Mechanism like e.g. the Elliot-Yafet coupling\cite{elliotyafet} has been suggested to cause spin-flip process with the
influence of impurities and phonons. \cite{koopmans1,koopmans2} However, Ref.~\onlinecite{bigot} argues against this, pointing to that experiments are not in favor of the suggested mechanism. Alternative mechanisms have instead been proposed, e.g. the super diffusive effect \cite{marco} involving spin angular momentum conserving super diffusion, where majority carriers have a high mean free path whereas minority carriers are less mobile. Very recently, Ma {\it et al.} \cite{ma} have proposed a dynamic spin-lattice-electron model to study demagnetization process of an iron thin film.

Although much effort has been devoted to the demagnetization process in laser induced femtosecond experiments, less attention has been paid to the aspect of the recovery of magnetism after a material has been subjected to a strong femtosecond pulse and cools off. In this letter, we report on a microscopical model for the remagnetization process, using first principles theory in combination with atomistic spin-dynamics simulations, and a three temperature model. We demonstrate that atomic magnetic moments as well as macroscopic magnetization evolve in time in a complex non-monotonous manner due to energy transfer between spin, lattice and electronic degrees of freedom for different starting values of electron temperatures.

In the initial stage of laser exposure to metals, the valence electrons absorb the energy from the laser while the lattice and spin remain in an unexcited state. Hence, a non-equilibrium situation occurs between the electron, lattice and spin subsystems. The electron-electron scattering takes place in less than 10 fs  or even faster, creating thermalized electrons. \cite{echenique} These electrons create spin-flip excitations across the Stoner gap, and hence reduce the size of the magnetic moment on an atomistic level, by single particle excitations (see e.g, Ref.~\onlinecite{mohn}). The size of the atomic moment decreases with an increase in temperature of the electronic sub-system,  and vanishes at a sufficiently high temperature of the electron gas, which we refer to as the Stoner-Curie temperature.\cite{mohn} Fermi-Dirac statistics is hence valid in the remagnetization process, at least after 10 fs, resulting in atomic magnetic moments that depend on the temperature of the electron sub-system. From self-consistent first principles theory it is actually possible to calculate the temperature dependence of the atomic moment as Gunnarson \cite{gunnar} did in the 70'ies, where Stoner-Curie temperatures were reported for iron ($\sim$6000 K) and cobalt ($\sim$5000 K).

A scenario we propose to describe the recovery of the magnetization in an intense laser excitation experiment is hence that initially the electron gas is excited to sufficiently high temperatures to significantly reduce the magnetic moment, and for significantly high temperatures, the atomic moment vanishes. The electron gas then cools down, as its energy dissipates in a way we describe below. As the temperature of the electron system is reduced below the Stoner-Curie temperature, atomic moments recover their magnitude as an equilibrium temperature is reached. During this time, the material is in a non-equilibrium situation in which the electron subsystem, the atomic moment subsystem and the lattice subsystem have separate temperatures, and hence separate dynamic evolution of their properties. Eventually the three subsystems reach the same temperature, something which can be described via the so called three temperature model, described below. \cite{Kampen}
 
In our work, we have calculated the time evolution of the electron temperature, using the three temperature model, and for each temperature of the electron subsystem we have calculated the atomic magnetic moments and the interatomic exchange interactions, using first principles density functional theory in which temperature effects enters via the Fermi-Dirac statistics. 
Hence the local moment for each atom is evolved with a thermal electron temperature, $T_{e}$, following the Fermi-Dirac statistics. The temperature dependent Fermi distribution function leads via the self-consistent cycle in density functional theory, to a change in the density of states(DOS), exchange splitting as well as the magnetic moment. The relaxation of the thermal electrons are determined by the electron-electron and electron-phonon collision rates, described below. 
The lattice is heated by electron-phonon interaction in a picosecond time scale. \cite{sven}

In parallel to the temporal evolution of the size of each atomic moment, we let the direction of each atomic magnetic moment evolve in time via the Landau-Lifshitz equation of motion. The basic philosophy behind this approach is similar to that of the Born-Oppenheimer approximation, that the atomic spin dynamics is significantly slower that the electron dynamics. \cite{antropov}
In our simulations, we hence combine the three temperature model (TTM) \cite{Kampen} with first principles theory, and atomistic spin dynamics (ASD) calculated by the UppASD software \cite{asd} using the Landau-Lifshitz-Gilbert (LLG) equation, where the ASD parameters are calculated from first principles theory. In ASD, the temporal evolution of individual atomic moments in an effective field, at a finite temperature of the spin-system, is governed by Langevin dynamics, represented through the stochastic differential equation of the LLG form,

\begin{eqnarray}
\frac{d{\bf m}_{i}(t)}{dt} &=& -\gamma{\bf m}_{i}(t)\times[{\bf B}_{i}(t)+{\bf b}_{i}(t)]-\\ \nonumber &&\gamma\frac{\alpha}{{\textit m}(t)}
{\bf m}_{i}(t)\times ({\bf m}_{i}(t)\times [{\bf B}_{i}(t)+{\bf b}_{i}(t)]),
\end{eqnarray} 
where ${\bf b}_{i}$ is a stochastic magnetic field involving thermal fluctuations, whose strength is defined as $D=\frac{\alpha}{1+\alpha^{2}}\frac{k_{B}T}{\gamma m}$. $\gamma$ is the gyromagnetic ratio, $\alpha$ is the damping parameter (we used $\alpha=0.1$ in our simulations) and $\bf{m}_{i}$ is an individual atomic moment on site $i$. $m$, $T$ and $k_{B}$ are magnitude of magnetic moment, temperature and Boltzmann constant respectively. Details of how LLG equation is evaluated in practice can be found 
in Ref.~\onlinecite{asd}. Compared to previous implementations of the Landau-Lifshitz equation there is a difference in the present work, in that both the direction and size of each atomic moment are allowed to change with time. This is indicated in Eqn.~1 by an explicit time dependence of the magnetic moment ${\bf m}_{i}(t)$, its size ${\textit m}(t)$ as well as the effective exchange field ${\bf B}_{i}(t)$. The direction of the effective field ${\bf B}_{i}(t)$ is determined by the exchange interaction between the atomic moment at site $i$ and all other moments, and is at each time-step given by $- \sum_j J_{ij} {\bf m}_j$. In our calculations we let the atomic moments initiate their time evolution from a collinear ferromagnetic configuration, as well as several canted non-collinear configurations. We observed very similar results for all starting configurations, and below we show only data from the collinear initial configuration. 

The spin temperature, $T_{s}(t)$, is an input temperature for the heat bath, and hence enters the stochastic field ${\bf b}_{i}(t)$. The ASD simulations were performed on a 20 x 20 x 20 bcc Fe system and a 20 x 20 x 20 hcp Co system, with periodic boundary conditions using calculated exchange parameters within the first ten coordination shells. No external field was applied in the simulations.

\begin{table}
\caption{The parameters used in the TTM model. For definitions see text.}
\begin{tabular}{|l||l||l|}
\hline
\hline
&    Fe    &    Co     \\
\hline
$ C_{l}[J{m}^{-3}{K}^{-1}]$ &2.2 x 10$^6$&2.07 x 10$^6$\\

$G_{ep}[J{(s{m}^{3}{K})}^{-1}]$ &4.05 x 10$^{18}$&4.05 x 10$^{18}$\\


$\gamma_{sp}[J{m}^{-3}{K}^{-2}]$ &670&662\\

$\tau_{M}$ &0.34 x 10$^{-12}$&0.34 x 10$^{-12}$\\
\hline
\hline
\end{tabular}
\end{table}
\begin{figure}[h]
\begin{center}
\includegraphics[scale=0.3]{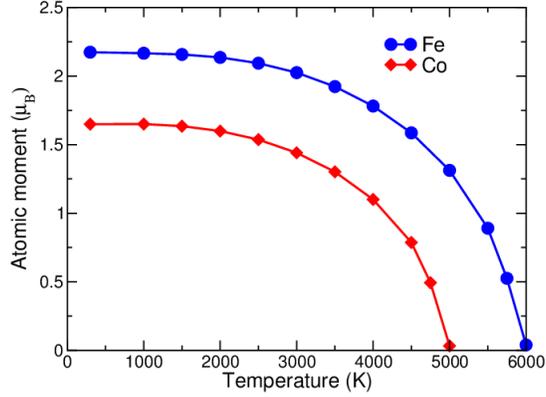}
\end{center}
\caption{(Color online) Temperature dependence of the local magnetic moment of Fe and Co, as calculated using density functional theory combined with temperature effects of the electron system, following the Fermi-Dirac statistics.}
\label{fig1}
\end{figure}

The exchange field $\bf{B}_{i}$ is determined by the interatomic exchange interactions in the Heisenberg model. The exchange parameters, and the atomic moments were calculated from first principles theory, as implemented in the exact muffin tin orbital (EMTO) method \cite{vitos} using the local force theorem. \cite{Liechtenstein} We adopted the local spin density approximation to the exchange correlation potential. In Fig.~1 we show for bcc Fe and hcp Co, the calculated atomic moments as a function of temperature. One observes that the Stoner-Curie temperatures for bcc Fe and hcp Co are around 6030K and 5000K respectively. It is curious to note that the Stoner-Curie temperature of Fe is larger than that of Co, which is opposite to the real Curie temperature of these two metals. Also, the Stoner-Curie temperature should reflect in some way the intra-atomic exchange (whereas the measured Curie temperatures of Fe and Co reflect primarily the inter-atomic exchange), which to a good approximation is given by the expression $IM^2/4$. In this expression, $I$ is the Stoner $I$ and $M$ is the size of the atomic spin moment. For Fe and Co the Stoner $I$ is calculated to be just over 0.7 eV for both elements \cite{brooks}, and since the atomic moment is larger for Fe than for Co, one would without electronic-structure effects, expect Fe to have the larger Stoner-Curie temperature. 

The calculated nearest and next nearest exchange interactions of bcc Fe and hcp Co are shown as a function of temperature in Fig.~2. We have evaluated interatomic exchange interactions up to ten shells, but Fig.~2 only shows the nearest and next nearest interactions, since they are dominant. It is clear from the figure that the nearest neighbor interaction stays ferromagnetic at all temperatures, and that its size decreases with increasing temperature. The next-nearest interaction is weaker for both systems, but has a very different behavior for Co and Fe. For hcp Co this interaction is almost temperature independent up to the Stoner-Curie temperature, where it disappears. For bcc Fe it decreases fast with temperature to become negative in a significant temperature interval before reaching the Stoner-Curie temperature. In this region, the magnetic configuration of bcc Fe is actually not ferromagnetic but non-collinear, as discussed also in Ref.~\onlinecite{lizarraga}.
 
\begin{figure}[h]
\begin{center}
\includegraphics[scale=0.3]{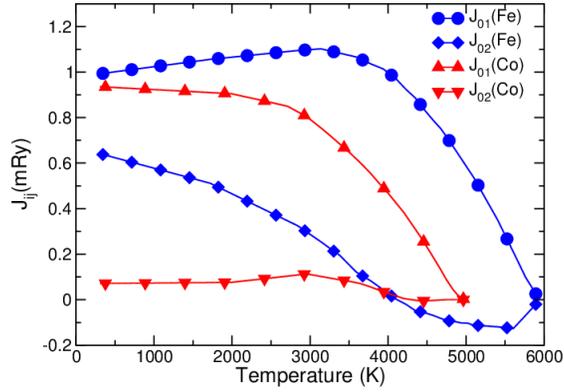}
\end{center}
\caption{(Color online) Calculated Heisenberg interatomic exchange parameters $ J_{ij}$  vs. temperature for bcc Fe and hcp Co. Both nearest and next nearest neighbor interaction parameters are shown.}
\label{fig2}
\end{figure}

Next we describe the details of the three temperature model (TTM), which we have used. The electron temperature $T_{e}(t)$, lattice temperature $T_{l}(t)$ and spin temperature $T_{s}(t)$ are coupled to each other within the TTM model in the form of three coupled differential equations,

$$C_{e}\frac{dT_{e}}{dt} = -G_{el}(T_{e}-T_{l})+P(t)-C_{e}\frac{(T_{e}-T_{room})}{{\tau}_{th}}$$
 $$ C_{l}\frac{dT_{l}}{dt}=G_{el}(T_{e}-T_{l})$$
$$\frac{dT_{s}(t)}{dt}=\tau^{-1}_{M}[T_{e}(t)-T_{s}(t)],$$
where $C_{e}$ and $C_{l}$ are the specific heats of the electron and lattice systems respectively. $C_{e}$ is defined as $C_{e}=\gamma_{sp}T$, where $\gamma_{sp}$ is the electronic specific heat constant. $G_{el}$ is the electron-phonon coupling constant which determines the rate of the energy exchange between the electron and lattice subsystems. The demagnetization process can be characterized by the demagnetization time $\tau_{M}$. In this model, a heat diffusion time ($\tau_{th}$) is added to the electron subsystem. This parameter determines the rate of energy dissipating from the material to reach the ambient temperature (room temperature). \cite{atxitia} Our solution to the three temperature model was obtained for the set of parameters \cite{param, koopmans}  listed in Table~1. 
 \begin{figure}[h]
\begin{center}
\includegraphics[scale=0.3]{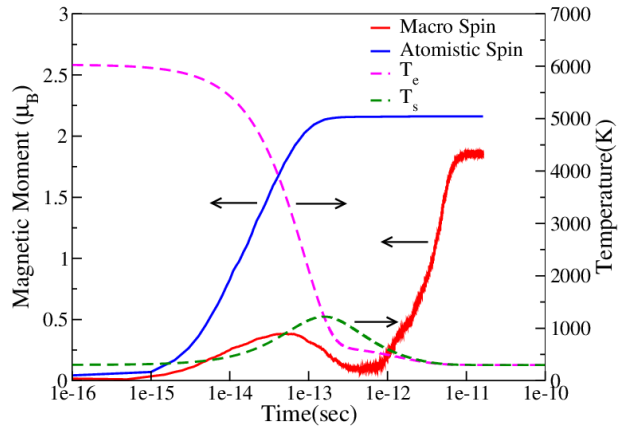}
\end{center}
  \caption{(Color online) Evolution of magnetic moment (per atom) of bcc Fe for an initial electron temperature of 6030K, with a heat dissipation time of 0.1 ps. The atomic as well as the macro spin moments are shown. The temperature of the electron and spin systems are also shown, with a scale on the right hand side of the figure.} 
\label{fig3}
\end{figure}

The simulations for bcc Fe are divided into three cases based on the initial value of the electron temperature, 1500 K, 3000 K and 6030 K. Further, the three cases are simulated using different heat diffusion times $\tau_{th}$ as, 60 ps, 1 ps and 0.1 ps. We describe in the Supplementary Information all the results for the three initial values of the electron temperature, and for the three different heat diffusion times. In the Supplementary Information we also show similar data for hcp Co, for three initial temperatures of the electron system 1500, 3000 and 5000 K and for heat diffusion times of 60 ps, 1 ps and 0.1 ps. Our calculations show that the most interesting results are found when the electron temperature is sufficiently high so that the atomic moment is significantly reduced from the ground state value. As an example of this behavior we show in Fig.~\ref{fig3} the case of bcc Fe when the initial electron temperature is the same as the Stoner-Curie temperature, and with 0.1 ps heat diffusion time. In this situation the temporal evolution of the atomic magnetic moment as well as the macroscopic magnetization has a particularly interesting behavior. As Fig.~\ref{fig3} shows, the atomic spin has a less dramatic behavior, increasing as the electron temperature cools down, and after $\sim$ 0.1 ps the atomic moment has reached its saturation value. The behavior of the macro spin is however much more interesting, where it increases to reach a local maximum just before 0.1 ps, which is followed by a decrease to reach almost a zero value after $\sim$ 0.5 ps. After this, the macro spin moment increases monotonically to reach its saturation after $\sim$ 10 ps. Similar to the finding of Beaurepaire {\it et al.} \cite{beaurepaire}, we observe a slight overshooting of the spin temperature before the electron temperature is cooled down to equilibrium.

The reason for this unexpected behavior can be understood from an inspection of the temperatures of the electron and spin systems. Initially the temperature of the spin system is low. As the temperature of the electron system is reduced, the atomic spins grow, however this happens without the atomic spins being subjected to thermal fluctuations, so that the atomic moments grow collinearly. However, after 0.05-0.1 ps, the temperature of the spin system has become significantly large to cause a disorder in the orientation of the atomic spins. As a matter of fact, in this time interval, the temperature of the spin system is above the Curie temperature and hence the macro spin vanishes. After sufficiently long time, the electron and spin systems equilibrate and reach the room temperature. The atomic moment and macro spin saturate in this case.

The simulated data shown in Fig.~3 are the most interesting, in the sense that the magnetization evolves in time in a highly non-monotonous behavior. Similar non-monotonic behaviors can also be found for different initial temperatures of the electron system, and for different heat diffusion times. This holds true both for bcc Fe and hcp Co, as discussed further in the Supplementary Information.
 
To summarize, the present study provides a microscopic picture of the remagnetization process of ultrafast pump-probe experiments that address magnetization dynamics. For the first time, we have combined LLG and first principles theories by taking into account the temporal evolution of the size of the atomistic moments as well as the exchange interactions, to allow for magnetization dynamics with both longitudinal and transverse fluctuations of the atomic moments. We performed the calculations up to 6030K for Fe and 5000K for Co, where the local moments are almost quenched to zero. Our results predict that in some cases, a highly non-intuitive behavior of the remagnetization can occur. The time-scales observed for the magnetization dynamics discussed here are well within the reach for experimental studies, and hopefully, our theoretical findings will motivate experimental works in this field. 
 
\section{Supplementary Information}
 In this supplementary information, we provide all the results of our simulations for Fe and Co for three different values of initial electron temperature and heat diffusion time (characterized by $\tau_{th}$).
For all the cases, we show the temporal evolution of magnetization (macro spin), local moment (atomistic spin), electron and spin temperatures.

\begin{figure}[h]

  \centering

  \begin{tabular}{ccc}


    \includegraphics[scale=0.2]{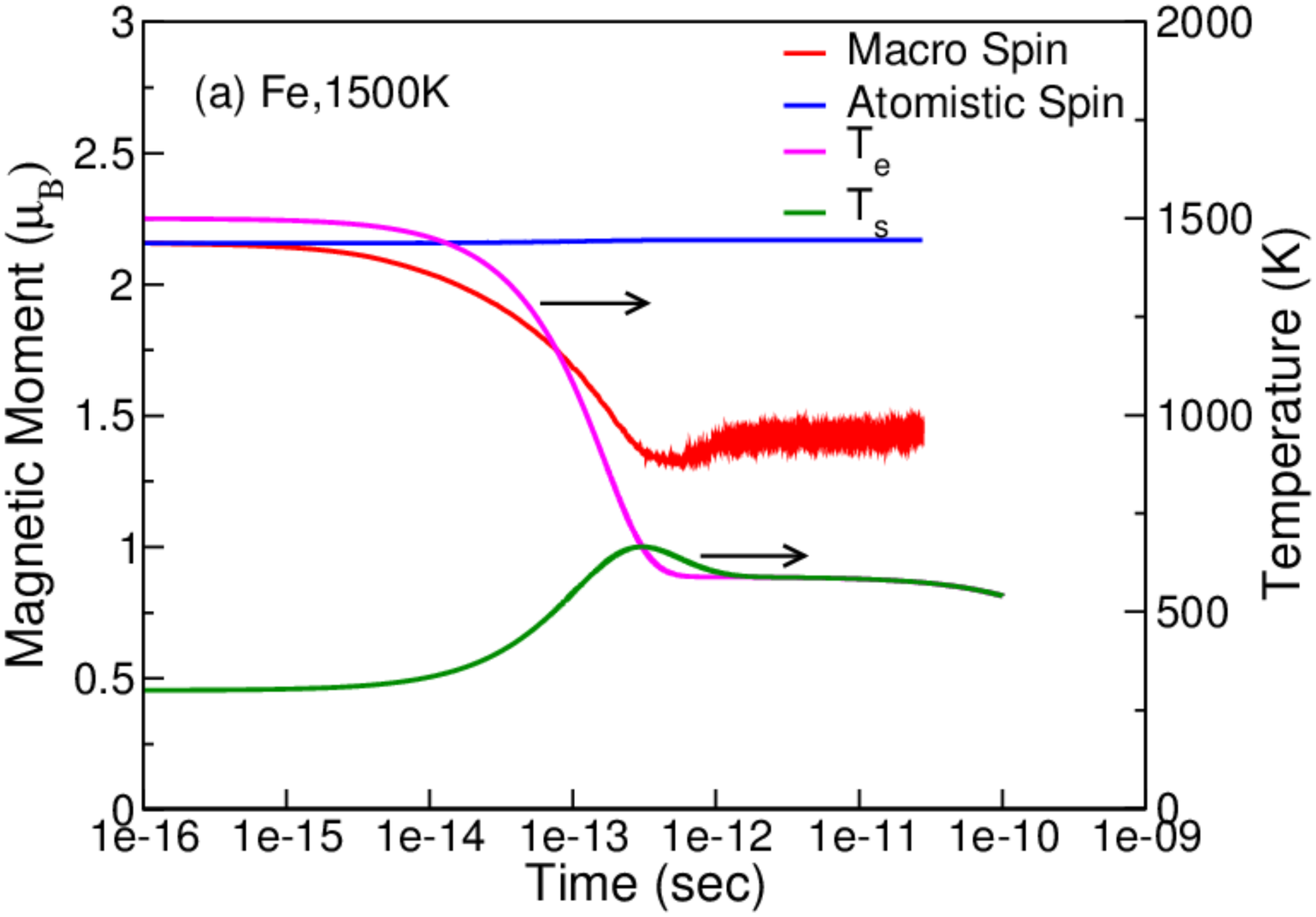}&

    \includegraphics[scale=0.2]{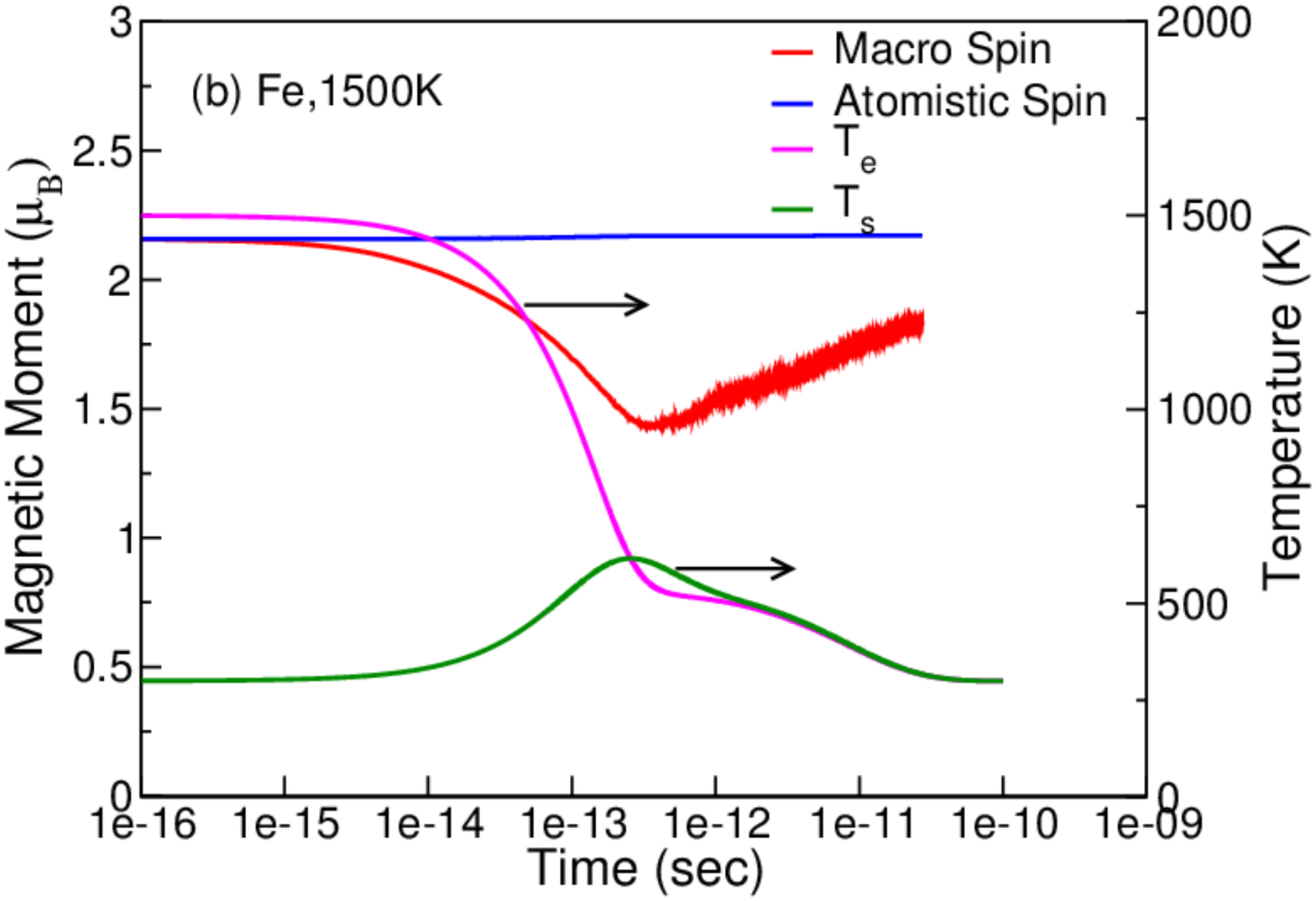}&
     \includegraphics[scale=0.2]{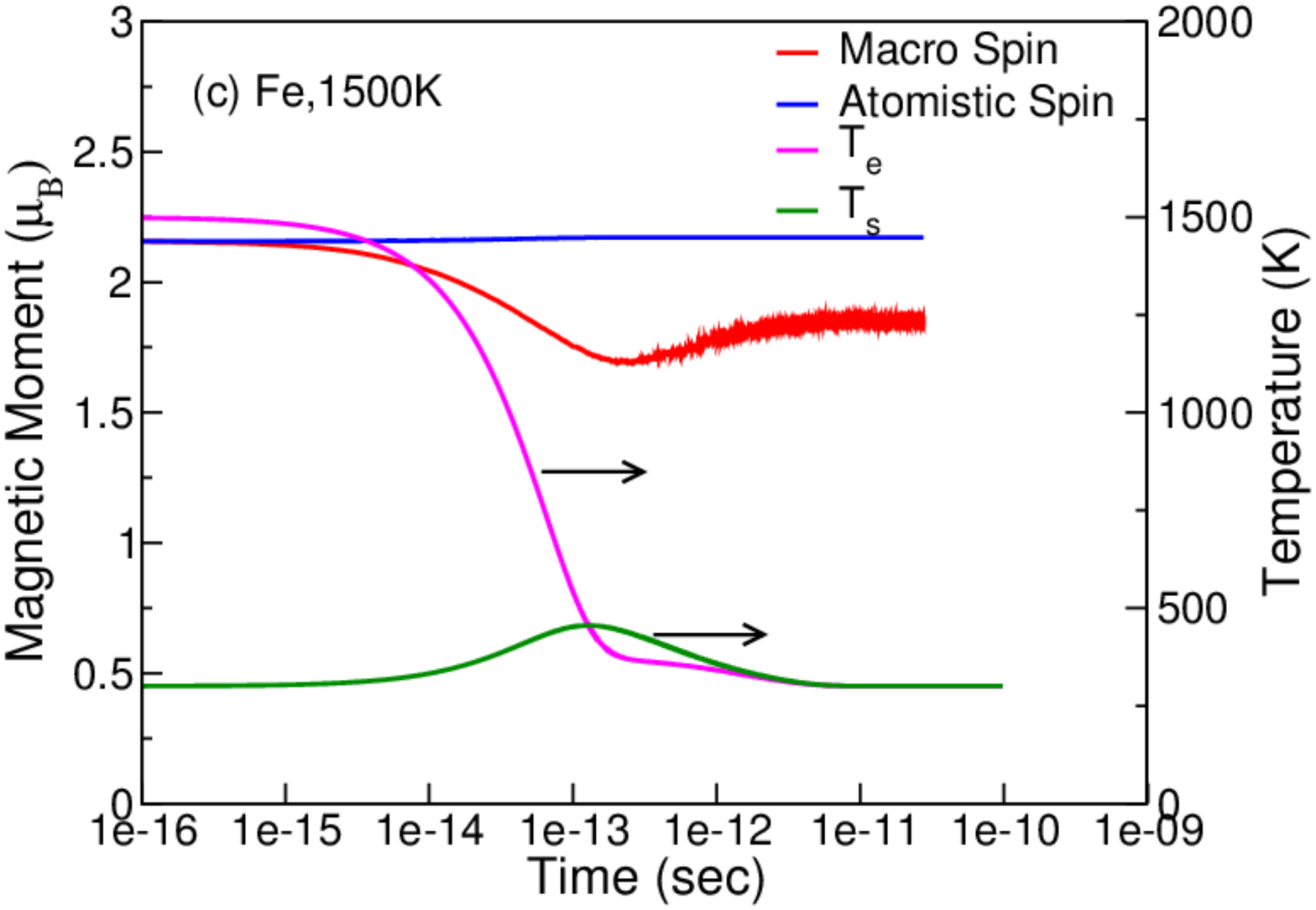} \\

     \includegraphics[scale=0.20]{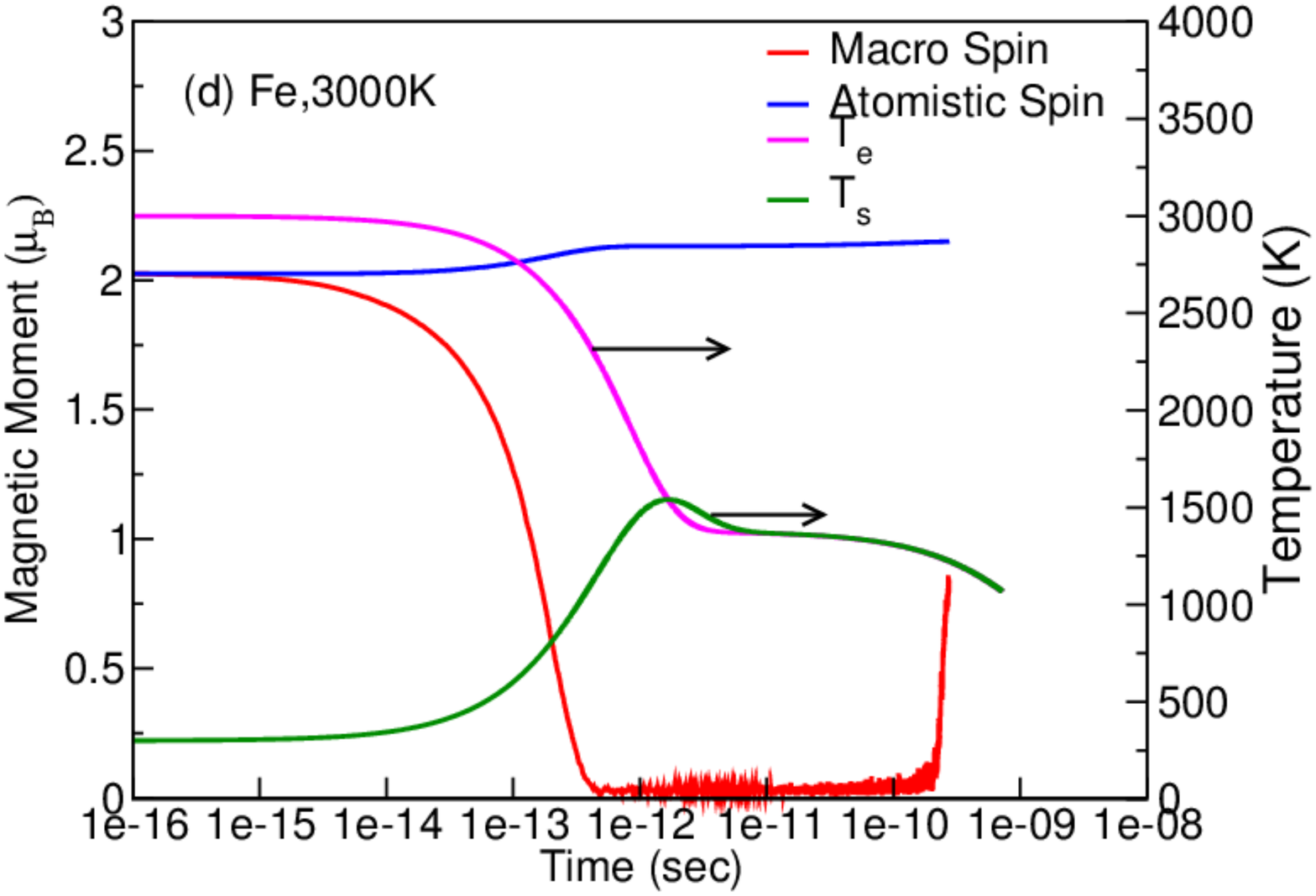}&

    \includegraphics[scale=0.20]{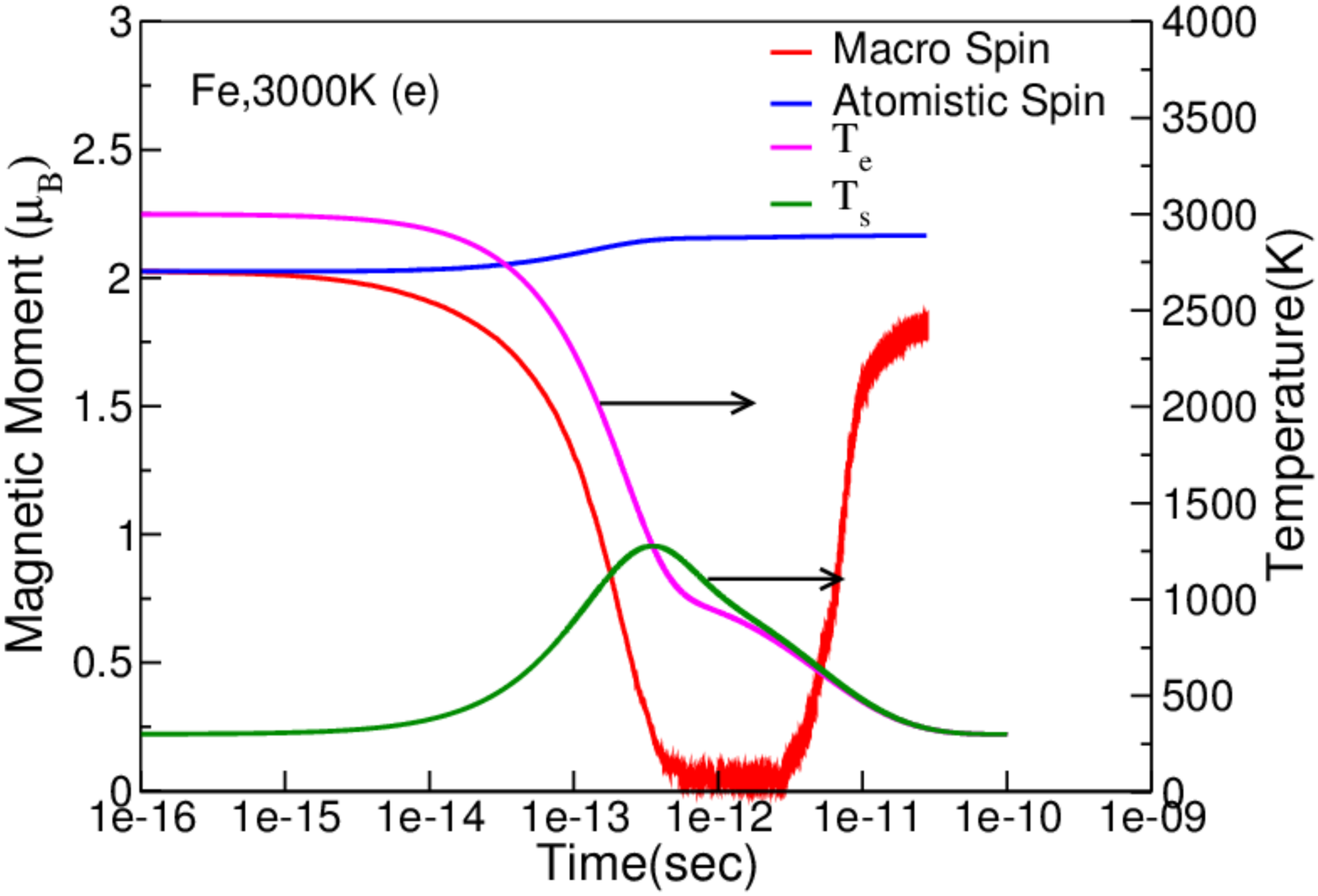}&
     \includegraphics[scale=0.20]{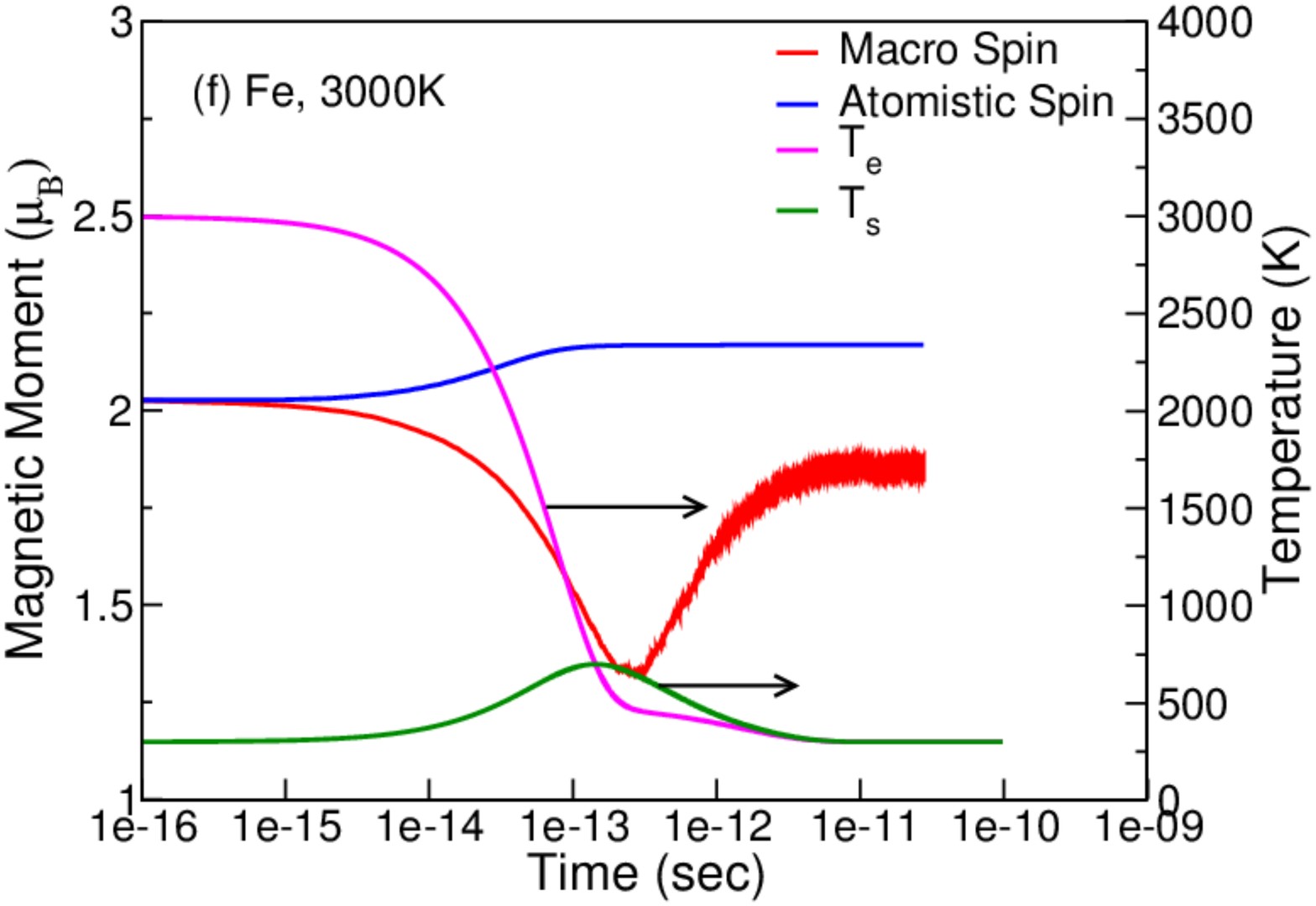}  \\

      \includegraphics[scale=0.20]{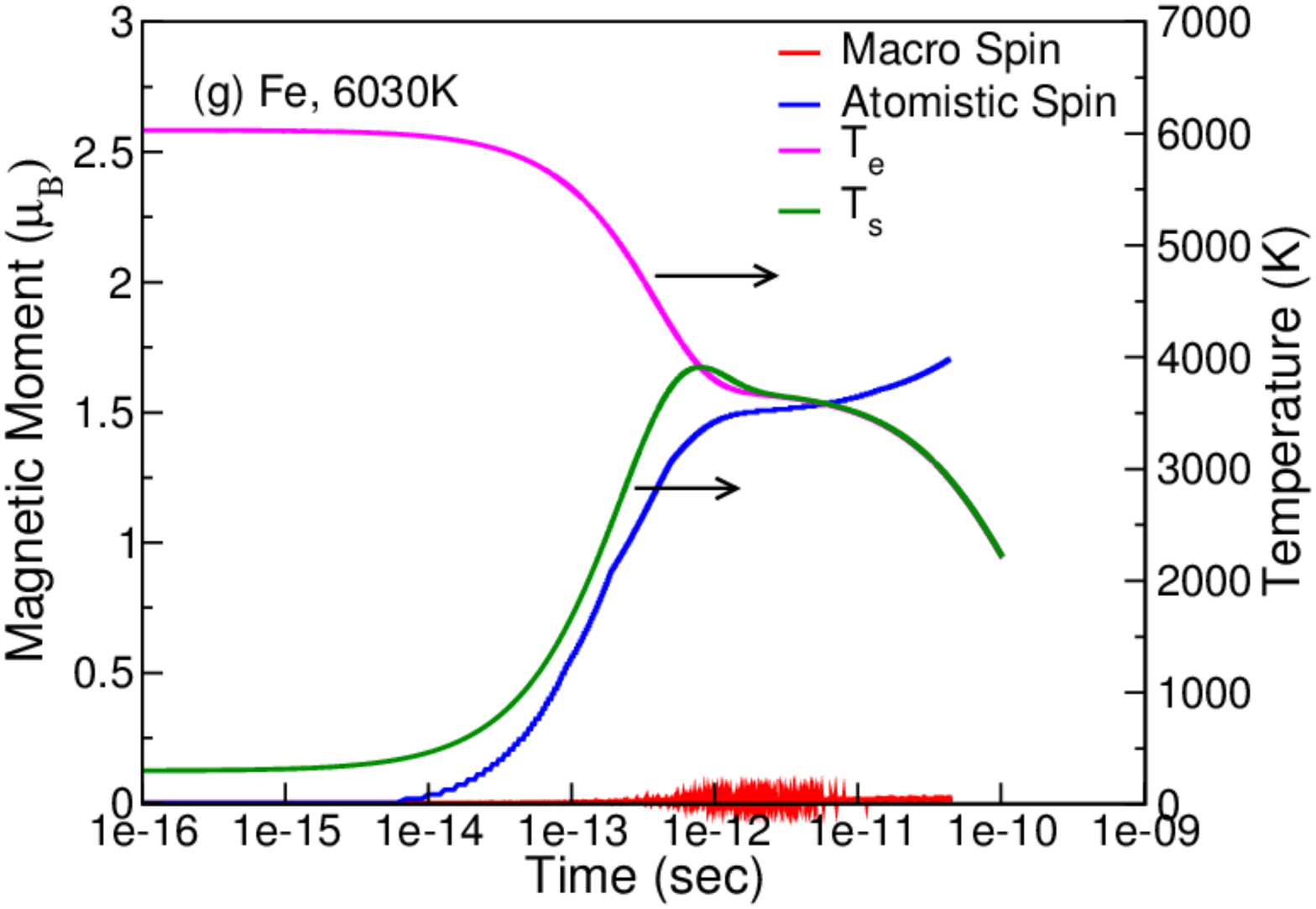}&

    \includegraphics[scale=0.20]{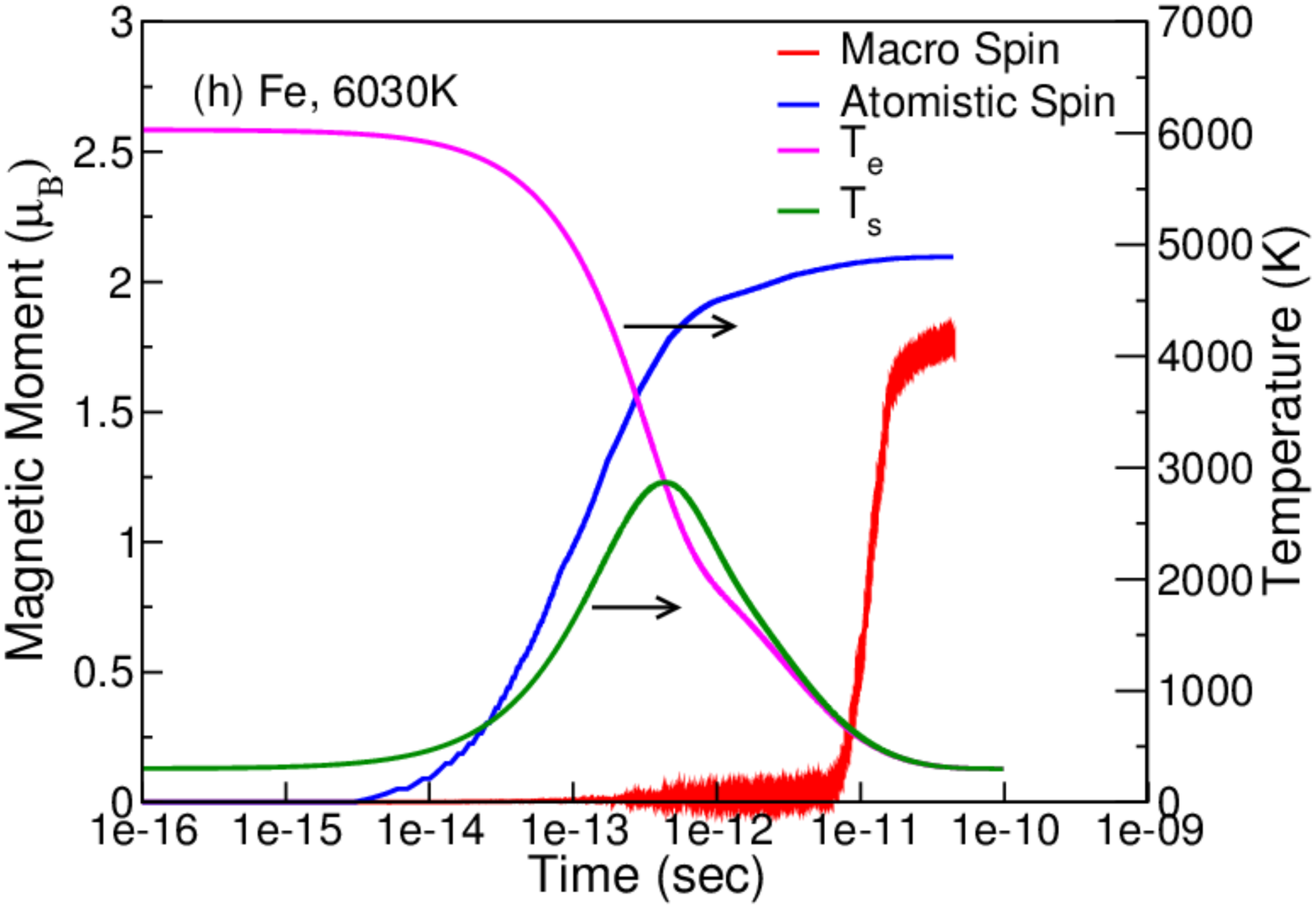}\\

 \end{tabular}
\caption{(Color online) Ultrafast magnetization dynamics for bcc Fe with macro spin M(t) (red), atomic moment  $\mu(t)$ (blue),  electron temperature $T_{e}$ (pink) and spin temperature $T_{s}$ (green). Dynamics shown for (a)-(c) $T_{e}=1500K$, $\tau_{th}=60, 1, 0.1 $ ps; (d)-(f) $T_{e}=3000K$, $\tau_{th}=60, 1, 0.1 $ ps; (g)-(h) $T_{e}=6030K$, $\tau_{th}=60, 1$ ps.}
  \label{SIfig1}
\end{figure}

In the first case, we consider the electron temperature of 1500K in bcc Fe, shown in Fig.~\ref{SIfig1}(a-c) with the heat diffusion time as mentioned in the main paper. It is observed that the demagnetization occurs within 0.5 ps as seen from the macro spin in Fig.~\ref{SIfig1}, more or less similar to what has been reported in previous theoretical works. \cite{beaurepaire,koopmans,refs}  One should note that slight deviations from previous works are expected as the initial condition is set at a higher electron temperature in our case and that our model is fundamentally different from previous works. In addition, we have used slightly different parameters in the three temperature model, a choice which was motivated by the experimental values. \cite{param} At 1500K, which is above the Curie temperature of bcc Fe, the local moments as well as the $J_{ij}$s are almost constant (evident from Fig.~1 and Fig.~2). Note that the spin temperature settles down at a value well below the Curie temperature so that a non-zero value of magnetization occurs at a sufficiently long time scale. This observation holds good for all the three electron temperatures studied except for $\tau_{th}$=60 ps at $T_{e}$=6030K, shown in Fig.~\ref{SIfig1}(g), where the electron- and spin-temperatures never become low enough to allow for a magnetically ordered state. Fig.~\ref{SIfig1} also shows that the behavior of the magnetization, i.e., the macro spin is different for different diffusion times. For $\tau_{th}$=1 and 60 ps, the magnetization does not reach saturation, for any initial value of the electron temperature, at least not in the time interval studied here.
\begin{figure}[h]

  \centering

  \begin{tabular}{ccc}


    \includegraphics[scale=0.20]{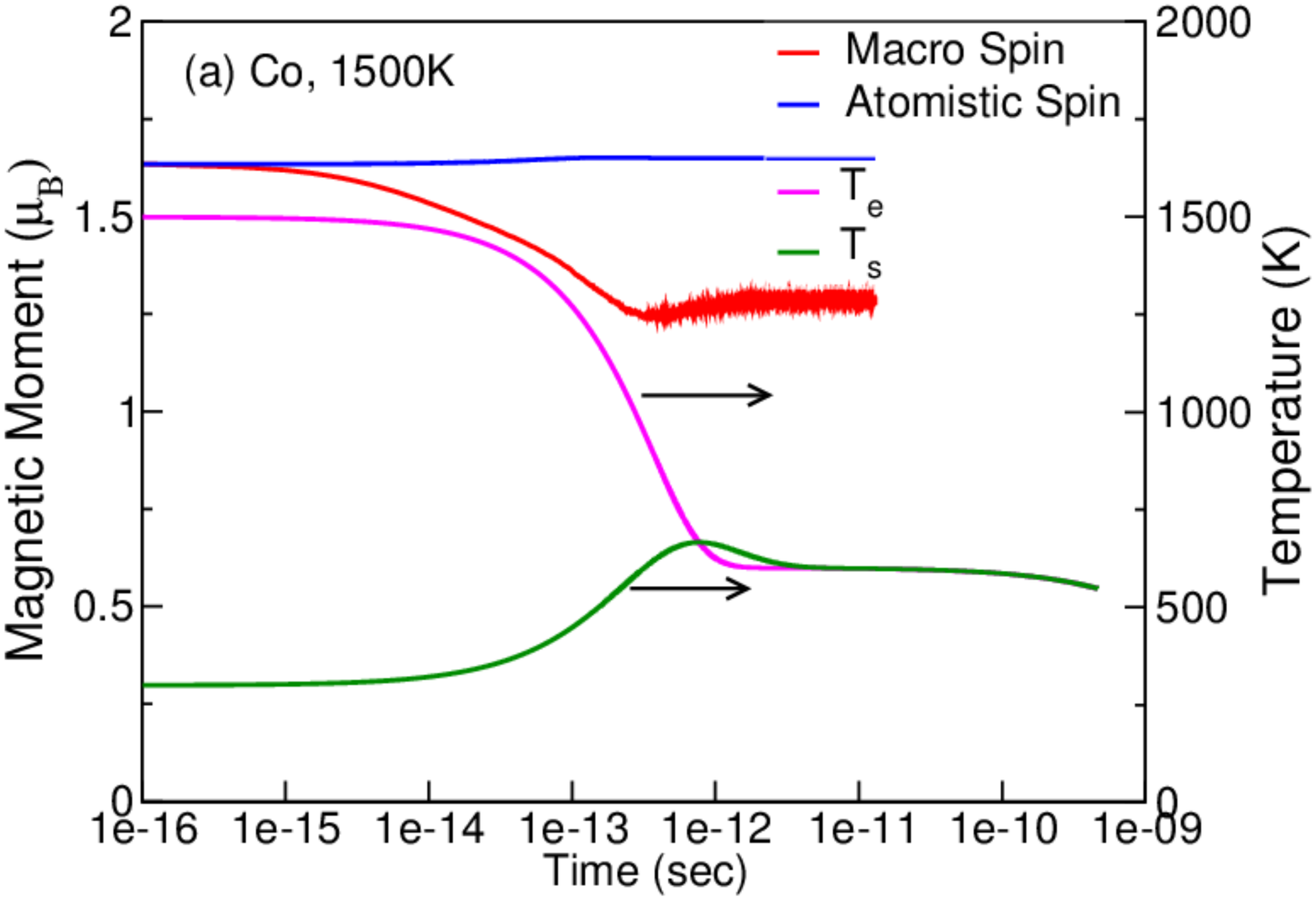}&

    \includegraphics[scale=0.20]{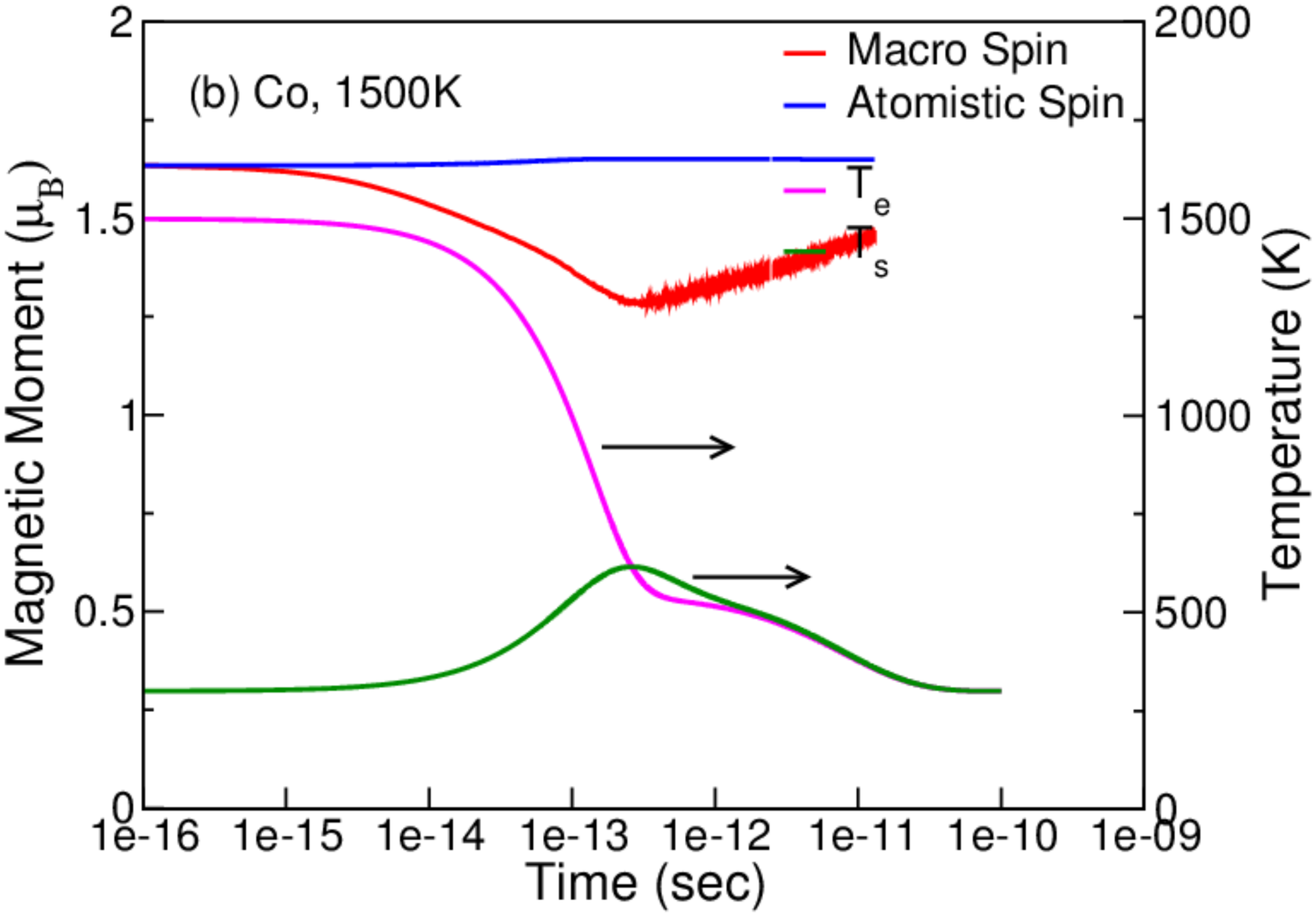}&
     \includegraphics[scale=0.20]{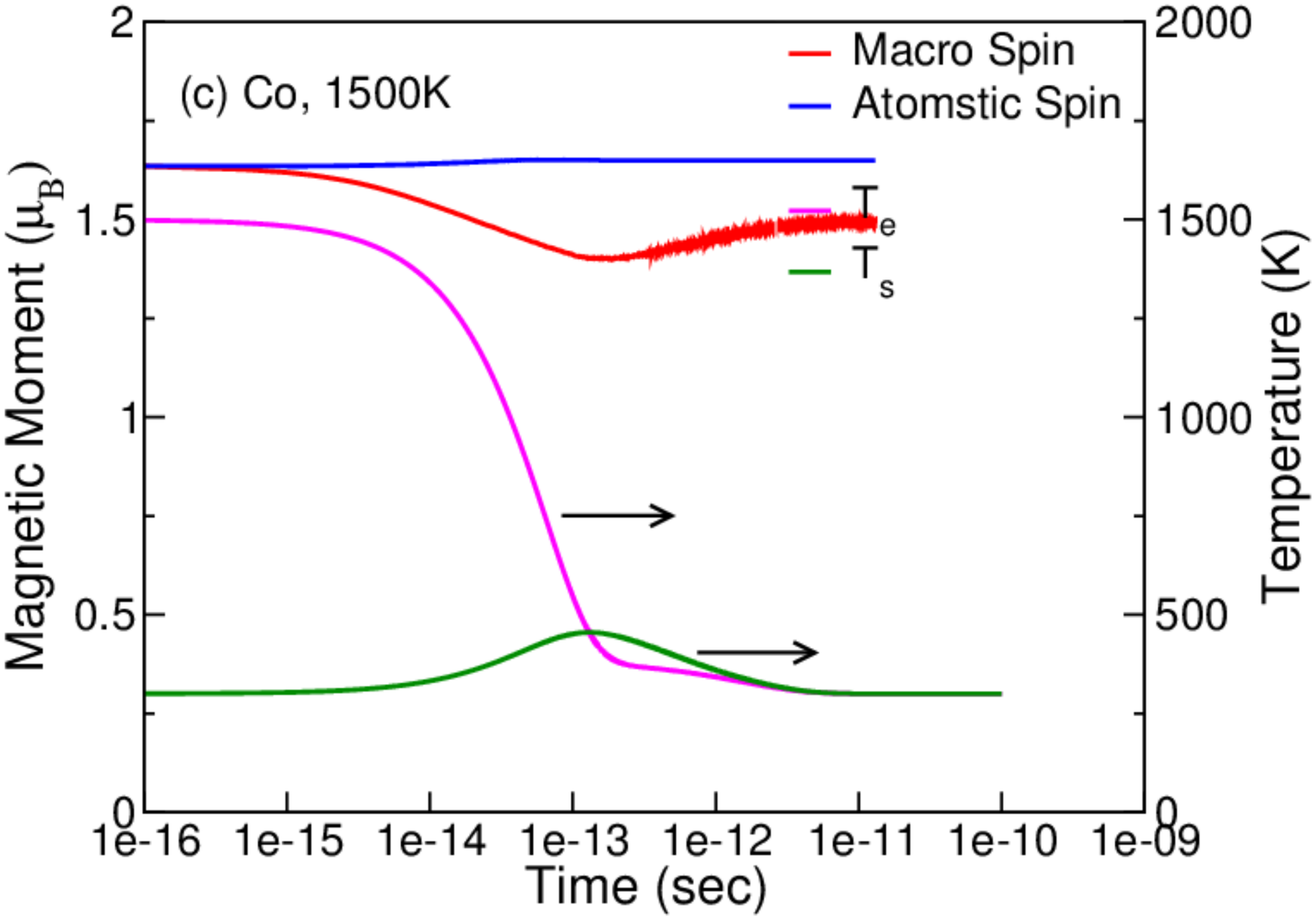}  \\

     \includegraphics[scale=0.20]{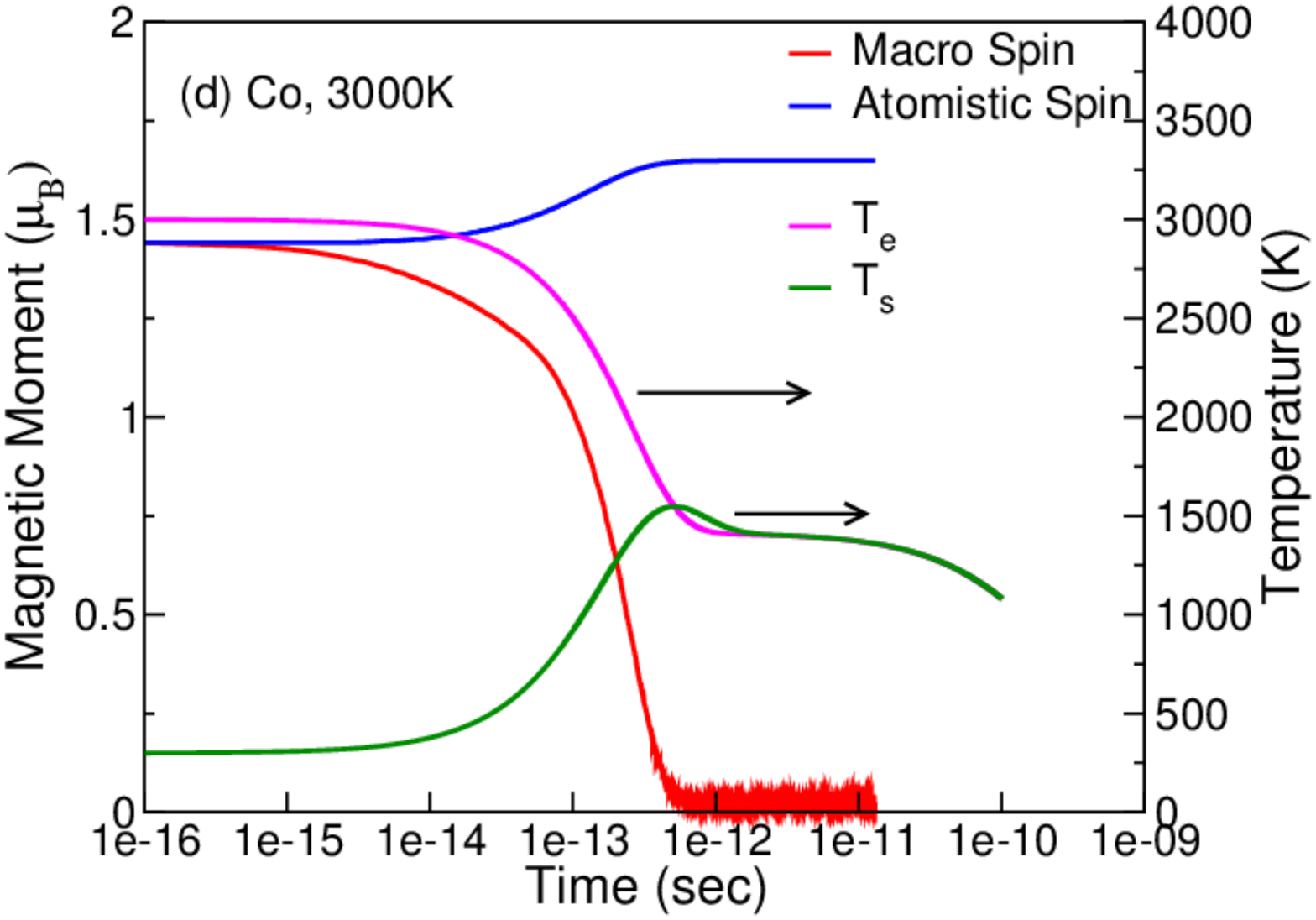}&

    \includegraphics[scale=0.20]{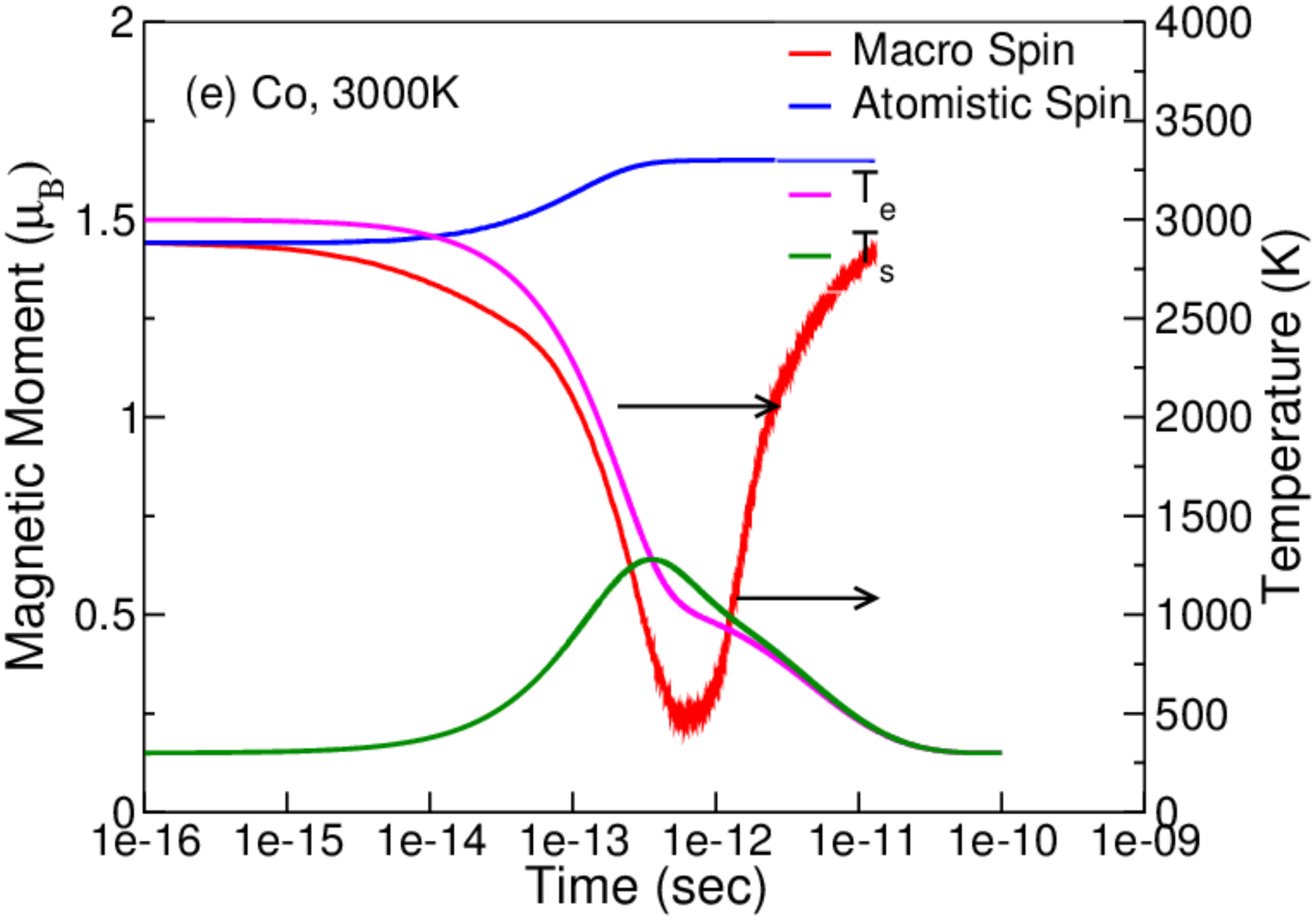}&
     \includegraphics[scale=0.20]{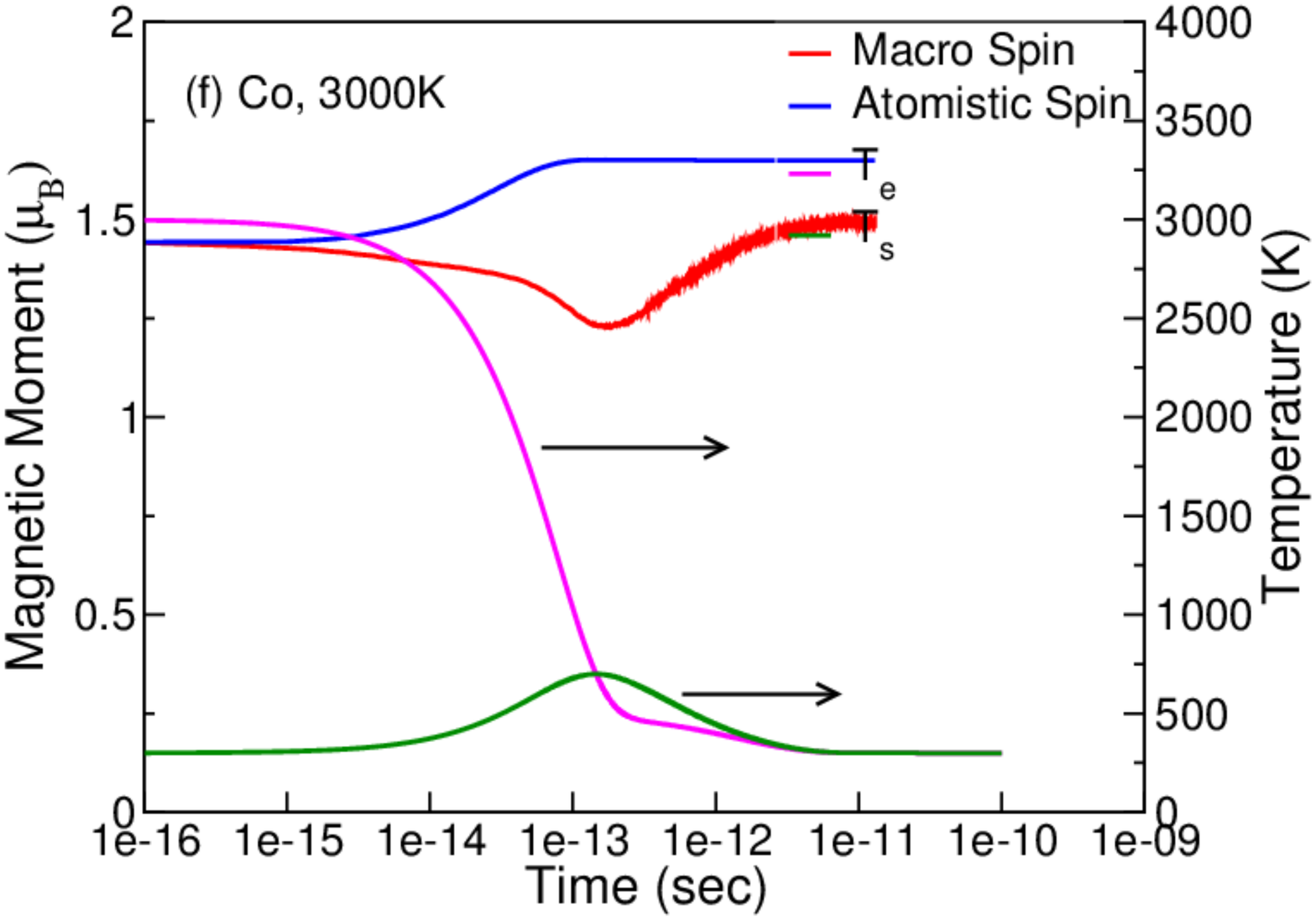}  \\

      \includegraphics[scale=0.20]{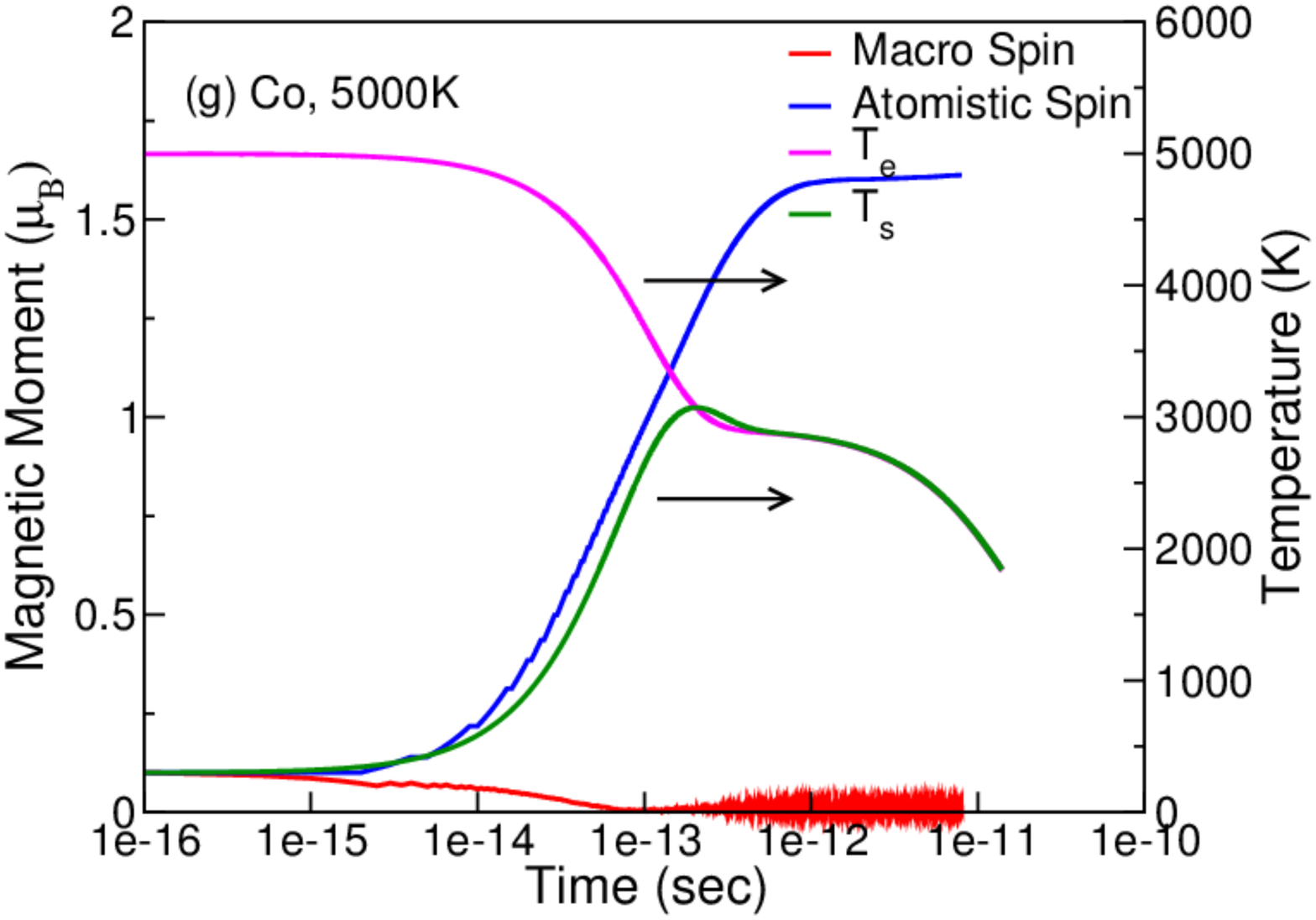}&

    \includegraphics[scale=0.20]{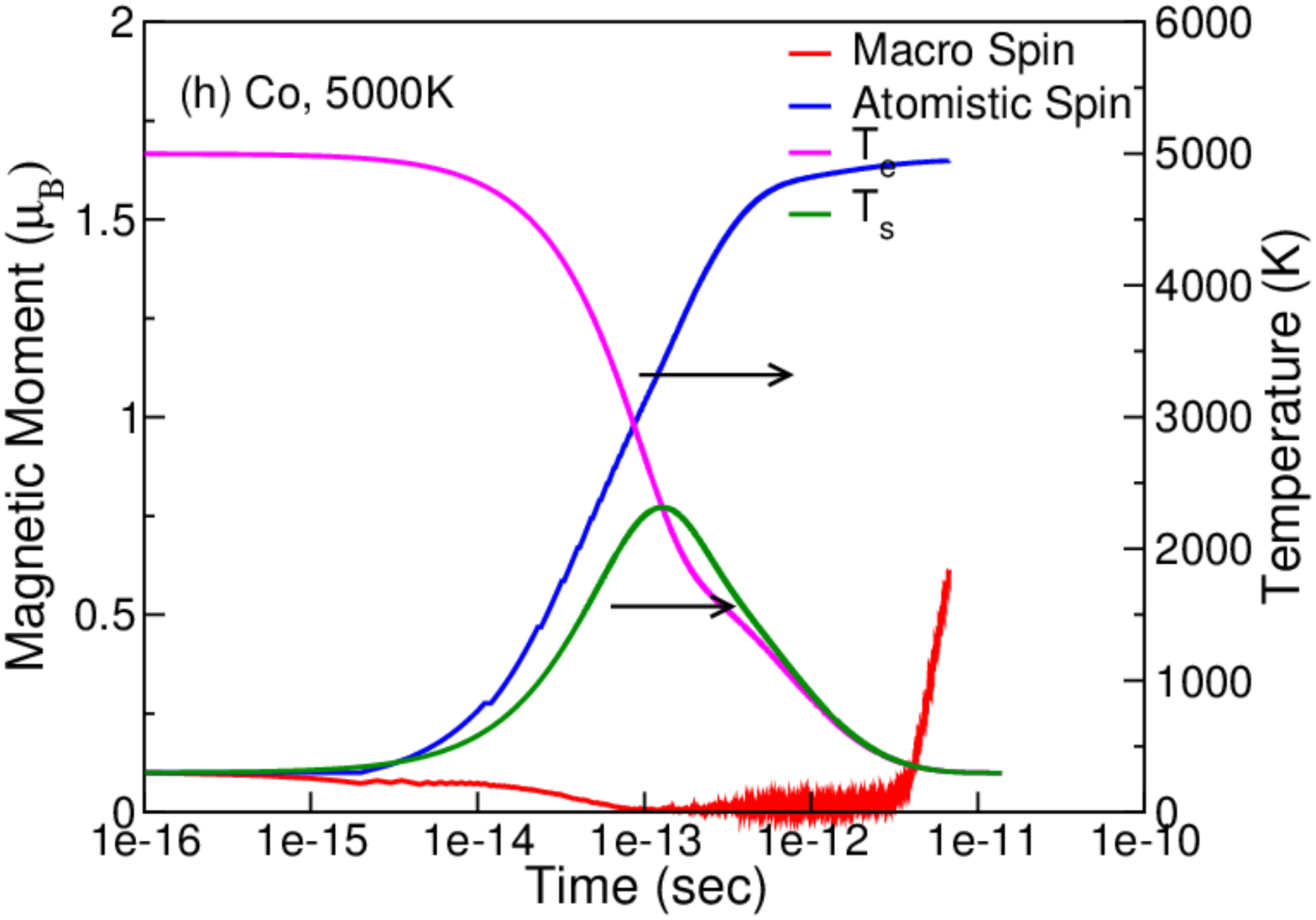}&
     \includegraphics[scale=0.20]{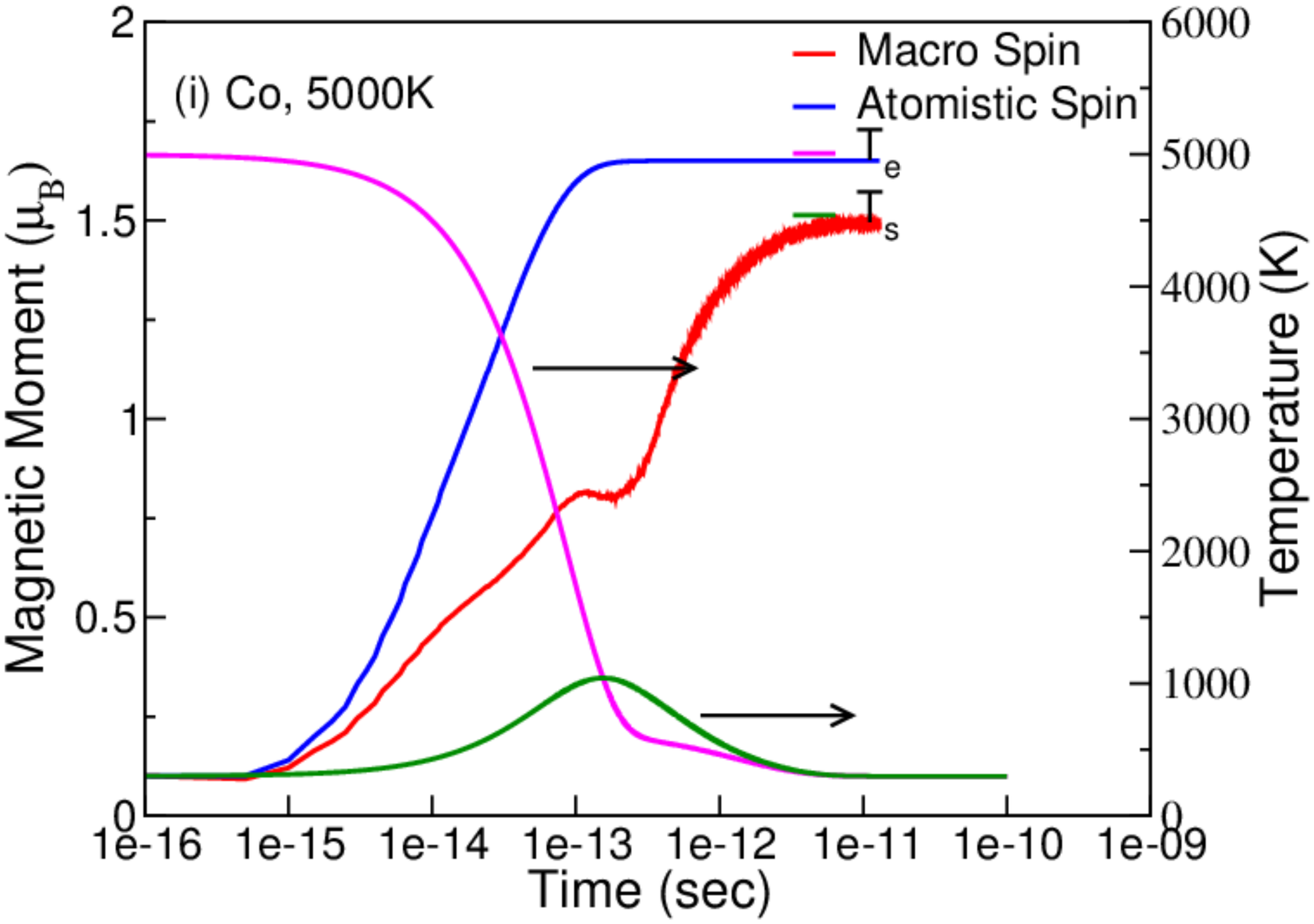}  \\

  \end{tabular}
\caption{(Color online) Ultrafast magnetization dynamics for hcp Co with macro spin M(t) (red), atomic moment $\mu(t)$ (blue),  electron temperature $T_{e}$ (pink) and spin temperature $T_{s}$ (green). Dynamics shown for (a)-(c) $T_{e}=1500K$, $\tau_{th}=60, 1, 0.1 $ ps; (d)-(f) $T_{e}=3000K$, $\tau_{th}=60, 1, 0.1 $ ps; (g)-(i) $T_{e}=5000K$, $\tau_{th}=60, 1, 0.1 $ ps.}
  \label{SIfig2}
\end{figure}

In the second case, the electron temperature was 3000K (shown in Fig.~\ref{SIfig1}(d) - (f)). From Fig.~\ref{SIfig1}, one can notice that the local moments start to drop rapidly around 3000K for Fe and Co.  Obviously, the value of the local moment is slightly smaller than the case of 1500K. In this situation, the macro spin is zero for a rather large time period, before becoming non-zero, especially when the diffusion time is 60 ps (Fig.~\ref{SIfig1}(d)). When the diffusion time is 1 ps (Fig.~\ref{SIfig1}(e)) the time interval for when the macro spin is zero is smaller, and finally when the diffusion time is 0.1 ps one notices only a dip in the macro spin curve (Fig.~\ref{SIfig1}(f)).

Finally, at the electron temperature of 6030K, the local moments are quenched to zero. The case for $\tau_{th}$=0.1 ps has already been shown in Fig.~3. Here we show the cases for $\tau_{th}$=60 and 1 ps. For heat diffusion time of 60 ps, remagnetization is not observed at all even for very long time scales. This is due to the fact that the spin temperature attains a value above 2000 K, which is much above the Curie temperature in the paramagnetic spin disordered phase. However, for smaller diffusion times, the remagnetization takes place but with a feature different from the cases of lower electron temperatures discussed above. 

We followed the same procedure to do the simulations for Co. The Stoner Curie temperature for Co is approximately 5000K, which is smaller than that of bcc Fe. At the same time, the Heisenberg exchange coupling parameters are stronger than those of Fe. The simulations shown in Fig.~\ref{SIfig2}(a-i) are for electron temperatures of 1500K, 3000K and 5000K and they have similar qualitative features as Fe. At the initial electron temperature of 1500K, the local Co moments remain essentially intact in size in the entire temporal range. This is similar to the case of Fe for the same value of electron temperature. At the initial electron temperature of 3000K, the atomic moments are slightly reduced initially in the simulation, but after sufficiently long time (pico seconds) a fully saturated atomic moment develops. 

At 5000K, the atomic moments are almost quenched to zero followed by an increase with descending value of electron temperatures. The most conspicuous case for Co is for $\tau_{th}$=0.1 ps, as shown in Fig.~\ref{SIfig2}(i), where at $\sim$ 5 ps the spin temperature is maximum and the magnetization curve attains as a result a shoulder, before increasing towards its saturation value. 

Finally, one may highlight some general observations on the time evolution of different physical properties for both Fe and Co. For all three values of electron temperatures, the spin temperatures settle down to a higher value than the initial one if the heat diffusion time is 60 ps. For lower values of $\tau_{th}$, the spin temperatures come down more or less to the same value of the spin temperature (300K) at the longest time considered. It is also observed that at a particular electron temperature, the maximum decrease in the value of magnetization occurs for the case with highest value of the diffusion time considered, i.e., 60 ps.
  
We gratefully acknowledge financial support from the Swedish
Research Council (VR). O.E. is in addition grateful to the ERC (project 247062 - ASD) and KAW foundation for support. Support from eSSENCE and SeRC are acknowledged.
We also acknowledge Swedish National Infrastructure for Computing (SNIC)
for the allocation of time in high performance supercomputers. We are grateful for fruitful and encouraging discussions with  C. Etz, J. Chico, M. Battiato, M.Pereiro and J. Fransson.








\end{document}